\documentclass[twocolumn,showpacs,preprintnumbers,amsmath,amssymb]{revtex4}
\usepackage{graphicx}% Include figure files
\usepackage{dcolumn}% Align table columns on decimal point
\def\bea{\begin{equation} }  
\def\eea{\end{equation} }  
\usepackage{bm}% bold math

\begin{document}

\title{Loop States in Lattice Gauge Theories}% Manuscript Title:\\with Forced Linebreak}% Force line breaks with \\

\author{Manu Mathur}
%\altaffiliation[TIFR, Mumbai]{%S.N. Bose National Centre for Basic Sciences\\
%JD Block Physics Department, XYZ University.}
\affiliation{% S. N. Bose National Centre for Basic Sciences} 
Tata Institute of Fundamental Research \\
Homi Bhabha Road, Mumbai 400 005, India \\ \\  
S. N. Bose National Centre for Basic 
Sciences\footnote{Permanent address, E.Mail: manu@bose.res.in}  \\ JD Block, Sector III, 
Salt Lake, Calcutta, India  
%% \textbackslash\textbackslash
}%
%%\author{Charlie Author}
%%%% \homepage{http://www.Second.institution.edu/~Charlie.Author}
%%\affiliation{
%%Second institution and/or address\\
%%This line break forced% with \\
%%}%

\date{\today}% It is always \today, today,

\begin{abstract}
We solve the Gauss law as well as the corresponding Mandelstam
constraints of $(d+1)$ dimensional SU(2) lattice gauge theory
in terms of harmonic oscillator prepotentials. This enables us
to explicitly construct a complete orthonormal and manifestly
gauge invariant basis in the physical Hilbert space. Further, we show
that this gauge invariant description represents networks
of unoriented loops carrying certain non-negative abelian fluxes 
created by the harmonic oscillator prepotentials.
The loop network is characterized by $3(d-1)$ gauge invariant
integers at every lattice site which is the number of physical degrees
of freedom. Time evolution involves local fluctuations of these loops. 
The loop Hamiltonian is derived. The generalization to SU(N) gauge group 
is discussed.
\end{abstract}

\pacs{11.15. Ha }

\maketitle

\section{Introduction} 

The idea that gauge theories should be formulated completely in terms of 
loops in space carrying electric fluxes is quite old, appealing and
has long history \cite{mans}. It is widely believed that 
QCD  written in terms of such gauge invariant colorless loops, instead of 
colored quarks and gluons, is better suited to study non-perturbative long 
distance physics like color confinement. Further, since the introduction 
of $SL(2,C)$ Yang Mills connections as the basic variables for 
gravity \cite{ash}, the loop formalism goes beyond the color invariant 
description of gauge theories and has a much wider reach. To this end, the 
Wilson loop approach, though geometrical and manifestly gauge invariant, 
suffers from the serious problem of over completeness due to Mandelstam 
constraints \cite{mans,rst}. Therefore, a most economical as well as 
complete description of gauge theories in d dimension in terms of gauge 
invariant loop 
states is an important issue and the subject of the present work. 
One would like to solve the Mandelstam constraints in the loop basis  
and study the loop dynamics without making any approximations or 
taking any particular limits \cite{rst}. We use the recently proposed 
prepotential 
formulation of lattice gauge theories \cite{manu} to explicitly construct an 
orthonormal and manifestly gauge invariant basis in the physical 
Hilbert space and thus solve the SU(2) Gauss law as well as the associated 
Mandelstam constraints. The ideas can be generalized to SU(N) group 
and discussed in the last section. In the SU(2) case, we show that the basis 
vectors describe networks of loops which carry positive integer abelian fluxes
created by the prepotential operators. Further, the action of the 
Hamiltonian on the loop basis too has simple interpretation of counting, 
creating and destroying the above abelian flux lines on the links. 
The loop network is characterized by $3(d-1)$ angular momentum quantum numbers 
at every lattice site which is the number of the physical degrees of freedom 
of the SU(2) theory. Therefore, this loop state description is  also a duality 
transformation \cite{sharat1,rob}  where the effect of compactness of the 
gauge group is contained in the discrete angular momentum quantum numbers 
labeling the loop states. In the simpler context of compact (2+1) and (3+1) 
U(1) gauge theories such duality transformations are known to isolate the 
topological magnetic monopole degrees of freedom leading to confinement 
\cite{pol}. The plan of the paper is as follows. After a brief introduction, 
we first construct the 
loop states in d=2 and study their dynamics. This keeps the discussion simple 
and also illustrates 
all the ideas involved. The corresponding analysis and results in arbitrary 
d dimension is then obvious and done next.  The generalization to SU(N) case 
is discussed at the end. 

We start with SU(2) lattice gauge theory in (d+1) dimension. The Hamiltonian 
is \cite{kogut}:  
\bea 
H=\sum_{n,i}tr E(n,i)^{2} +K\sum_{plaquettes}tr \Big(U_{plaquette} + h.c\Big).     
\label{ham}   
\eea 
where K is the coupling constant. The index n labels the site of a d-dimensional
spatial lattice and $i,j (=1,2,...d)$ denote the unit vectors along the links.
Each link $(n, i)$ is associated with a symmetric top whose configuration,
i.e the rotation matrix from space fixed to body fixed frame, is given by the
operator valued SU(2) matrix $U(n,i)$. The angular momenta with respect 
to space fixed and the body fixed  frames are denoted by 
$E_{L}^{a}(n,i)$ and $E_{R}^{a}(n+i,i)$. More explicitly, 
$E_{L}(n,i)$ and $E_{R}(n+i,i)$ generate the gauge transformations 
at the lattice sites  n  and  n+i  respectively. 
They commute with each other and satisfy: 
$E_{L}(n,i).E_{L}(n,i) = E_{R}(n+i,i).E_{R}(n+i,i) \equiv E(n,i).E(n,i)$ 
as the total angular momentum is same in both the frames. 
Therefore, a complete basis at every 
link (n,i) is given by $|j(n,i),m(n,i),\tilde{m}(n,i)>$ where $j(n,i),m(n,i),
\tilde{m}(n,i)$  are the eigenvalues of $E(n,i).E(n,i), E_{L}^{3}(n,i)$ 
and $E_{R}^{3}(n+i,i)$ respectively. We now exploit the Schwinger boson 
representation of the angular momentum algebra \cite{schwinger} to 
define harmonic oscillator prepotentials on the links: 
\begin{eqnarray}
\label{sb}
E_{L}^{a}(n,i) &  \equiv &  a^{\dagger}(n,i)\frac{\sigma^{a}i}{2} a(n,i); \\ 
E_{R}^{a}(n+i,i) & \equiv & b^{\dagger}(n+i,i)\frac{\sigma^{a}}{2}b(n+i,i). 
\nonumber 
\end{eqnarray}
The gauge transformation properties of the angular momenta, 
$E_{L}(n, i) \rightarrow \Lambda(n) E_{L}(n, i) \Lambda^{\dagger}(n)$ and  
$E_{R}(n+i, i) \rightarrow \Lambda(n+i) E_{R}(n+i, i) \Lambda^{\dagger}(n+i)$, 
imply that the Schwinger bosons belong to the fundamental 
representations of the the gauge group, i.e: 
\bea
a^{\dagger}_{\alpha}(n,i)\rightarrow\Lambda(n)_{\alpha\beta} 
a^{\dagger}_{\beta}(n,i); b^{\dagger}_{\alpha}(n,i)\rightarrow\Lambda(n)_{\alpha\beta} b^{\dagger}_{\beta}(n,i). 
\label{gt3}
\eea
Therefore, $a^{\dagger}_{\alpha}(n,i)$ (left oscillator) and $b^{\dagger}_{\alpha}(n+i,i)$ 
(right oscillator) create spin half charges  
at left and right ends of the link (n,i) respectively. The total angular momentum 
being same in both the frames implies: 
\bea
a^{\dagger}(n,i).a(n,i) = b^{\dagger}(n+i,i).b(n+i,i)  \equiv N(n,i).  
\label{consho} 
\eea 
Thus,  besides SU(2) gauge invariance (\ref{gt3}) at every lattice site, we get 
an addition abelian gauge invariance on every lattice link:  
\begin{eqnarray} 
\label{u1}
a^{\dagger}_{\alpha}(n,i) & \rightarrow & \left(expi\theta(n,i)\right) ~ 
a^{\dagger}_{\alpha}(n,i);   \\ 
b^{\dagger}_{\alpha}(n+i,i) & \rightarrow & \left(exp-i\theta(n,i)\right) ~
b^{\dagger}_{\alpha}(n+i,i) \nonumber.   
\end{eqnarray} 
So we have gone from the electric field (or angular momentum), link operator 
description of  Kogut-Susskind Hamiltonian (\ref{ham}) to an equivalent description 
of SU(2) lattice gauge theory which is in terms of harmonic oscillator prepotentials
with $SU(2) \otimes U(1)$ gauge invariance. With the simple $SU(2) \otimes U(1)$  
gauge transformations (\ref{gt3}) and (\ref{u1}), we are well equipped to construct 
explicitly a manifestly gauge invariant and orthonormal loop basis. 
 
\section{The Loop States in d=2} 

\begin{figure}[t]
\begin{center}
\includegraphics[width=0.3\textwidth,height=0.3\textwidth]
{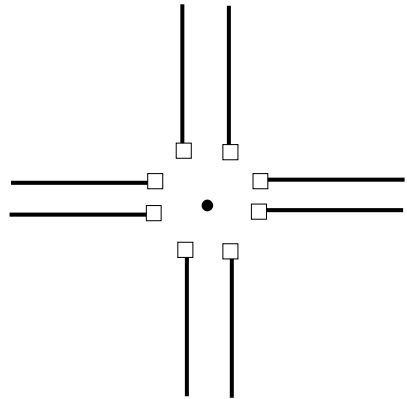}
\end{center}
\vspace{-5mm}
\caption{A graphical representation of the  $SU(2) \otimes U(1)$  
gauge transformations in d=2 at site n with $j_{1}=j_{2}=j_{3}=j_{4}=1$. 
The abelian flux lines graphically solves the abelian Gauss law 
(\ref{consho}). The (Young tableau) boxes at the ends of abelian flux 
lines represent the fundamental (spin 1/2) representation of SU(2) 
which acts at the lattice site n.}
\label{fig:figure1}
\end{figure}
The U(1) gauge invariance (\ref{u1}) and it's Gauss law (\ref{consho}) 
simply state that the number of the left ($a^{\dagger}(ni)$) and the right 
($b^{\dagger}(n+i,i)$) oscillators is the same on any link. We denote this 
integer number by $2j(n,i)$. The abelian Gauss law constraints are solved 
easily by drawing  $2j(n,i)$ lines on every link (n,i). 
Each of these $2j(n,i)$ lines represents the U(1) charge (= $+ 1$) of 
$a^{\dagger}(n,i)$ (see (\ref{u1})) and henceforth will be called abelian charge 
or abelian flux line. To illustrate, a simple example with 
$j(n,1)=j(n,2)=j(n-1,1)=j(n-2,2) =1$ is shown in Figure 1.  With U(1) Gauss 
law satisfied, we now deal with SU(2) gauge invariance. 
Under SU(2) gauge transformations (\ref{gt3}), we note that 
$a^{\dagger}(n,i)$ [located at the starting point of the link 
(n,i)] and $b^{\dagger}(n,i)$ [located at the end point of the 
link (n-i,i)] transform together by $\Lambda(n)$. Therefore, it is convenient 
to group them together and define $a^{\dagger}[n,i]$ with i=1,2,3,4. 
More explicitly, $a^{\dagger}[n,1] \equiv a^{\dagger}(n,1), 
a^{\dagger}[n,2] \equiv a^{\dagger}(n,2), a^{\dagger}[n,3] 
\equiv b^{\dagger}(n,1), a^{\dagger}[n,4] \equiv b^{\dagger}(n,2)$. 
Therefore, each of the four $a^{\dagger}[n,i]$ can be represented by 
a Young tableau (YT) box belonging to the SU(2) group which acts 
at the site n. 
Thus, to get SU(2) gauge invariance,  we have to construct all possible 
spin singlets out of $\left(\sum_{i=1}^{2d=4} 2j[n,i]\right)$  YT boxes. 
This is a simple problem: all possible spin zero  operators are of the form 
$a^{\dagger}[n,i].\tilde{a}^{\dagger}[n,j] \equiv 
\epsilon_{\alpha\beta}a^{\dagger}_{\alpha}[n,i]
a^{\dagger}_{\beta}[n,j]$, where $\epsilon_{\alpha\beta}$ is the completely 
antisymmetric tensor and corresponds to putting two boxes of type i and j 
in a vertical column.  In Figure 1, we represent this by linking 
a line of type i with a line of type j ($i \neq j$) (see Figure 2).  
Thus for SU(2) gauge invariance all abelian 
flux lines must be mutually linked and no self linking is allowed. 
Therefore, the necessary condition on the number of abelian flux lines 
on the links [n,i] to give SU(2) gauge invariant state(s) at site n is: 
\begin{eqnarray} 
2j(n,i) = \sum_{j\neq i} l_{ij},~~ l_{ij}=l_{ji},~~ l_{ij} \in {\cal Z_{+}},  
\label{part} 
\end{eqnarray}
where ${\cal Z_{+}}$ denotes the set of all non-negative integers and 
$l_{ij}$ are the linking numbers amongst $i$ and $j$ types of abelian 
flux lines. The partition (\ref{part}) represents the manifestly SU(2) 
gauge invariant state: 
\begin{eqnarray}
|\vec{l}> = \prod_{{}^{{i},{j}=1}_{~{j} > {i}}}^{2d = 4} 
\left(a^{\dagger}[i].\tilde{a}^{\dagger}[j]\right)^{l_{{i}j}}
|0> . 
\label{giv2} 
\end{eqnarray} 
The pattern $|l_{12}(n),l_{13}(n),l_{14}(n),l_{23}(n),l_{24}(n),l_{34}(n)>$ 
will be used to characterize the states in (\ref{giv2}).  
Thus given any network of loops on the lattice we can construct a 
manifestly $SU(2) \otimes U(1)$ gauge invariant basis (\ref{giv2}) 
characterized by $d(2d-1)$ integer quantum numbers at every lattice 
site. The disadvantage of the basis (\ref{giv2}) is that (like Wilson 
Loop basis) it is not orthonormal and it is over complete. 
To show this, we consider three distinct basis vectors contained 
in the set (\ref{giv2}): $|\vec{l}_1> = |100001>,  
|\vec{l}_2> = |010010>$ and $|\vec{l}_3> =|001100>$. We find 
that they are linearly related: $|\vec{l}_1> = |\vec{l}_2> - 
|\vec{l}_3>$ due to the identity: 
\begin{eqnarray}   
\left(a^{\dagger}[1].\tilde{a}^{\dagger}[2]\right) 
\left(a^{\dagger}[3].\tilde{a}^{\dagger}[4]\right)  
\equiv  \left(a^{\dagger}[1].\tilde{a}^{\dagger}[3]\right)
\left(a^{\dagger}[2].\tilde{a}^{\dagger}[4]\right) \nonumber \\ 
- \left(a^{\dagger}[1].\tilde{a}^{\dagger}[4]\right)
\left(a^{\dagger}[2].\tilde{a}^{\dagger}[3]\right).  ~~~ 
\label{mandi}
\end{eqnarray} 
Infact, the identity (\ref{mandi}) is the basic SU(2) Mandelstam 
identity written in terms of prepotentials. To solve the problem 
of over-completeness, we notice that the states (\ref{giv2}) obtained 
by different possible contractions of the abelian flux lines are all 
characterized by $|J[n,i].J[n,i] = j[n,i](j[n,i]+1), J 
\equiv J^{a}_{total} =0>$. However, the intermediate angular momentum 
quantum number labels are missing \cite{sharat1}. 
Therefore, we choose the following angular momentum addition scheme: 
$J[1]+J[2] \rightarrow J[12]+J[3] \rightarrow J[(12)3]+ J[4]=J  \equiv 0$ 
and label the the common eigenvectors by the corresponding eigenvalues: 
$|j_{1},j_{2},j_{12},j_{3},j_{123}=j_{4},j=j_{total}=0> \equiv 
|j_{1},j_{2},j_{12},j_{3},j_{4}>$. Thus, we get the (missing) operator 
$(J[n,1]+J[n,2])^{2}$ in this scheme which is  yet to be diagonalized in the basis 
(\ref{giv2}) with eigenvalue $j_{12}$. This diagonalization problem is again 
simple: after linking $l_{12}$ boxes from $2j_{1}$ YT boxes on the 
link (n,1) with $l_{12}$ boxes from $2j_{2}$ YT boxes on the link (n,2), 
we should be left with $2j_{12}$ boxes which are not linked (and symmetrized). 
Therefore, $l_{12}=j_{1}+j_{2}-j _{12}$. As total angular 
momentum is zero this also fixes $l_{34}=j_{3}+j_{4}-j _{12}$. 
This, in the example of Figure 1, is illustrated in Figure 2.  
The final orthonormal and manifestly SU(2) gauge invariant states are: 
\begin{eqnarray}
\label{fr}
|j_{1},j_{2},j_{12},(j_3),j_{123}=(j_{4})>  \equiv |j_1,j_2,j_{12} \rangle 
~~~~ ~~~~~~\\
=  N(j) \sum_{{}^{l_{13},l_{14}}_{l_{23},l_{24}}}\hspace{-0.05cm}
{}^{{}^\prime} 
\prod_{{}^{i,j}_{i < j}} \big((l_{ij})!\big)^{-1} \big(a^{\dagger}[n,i].
\tilde{a}^{\dagger}[n,j]\big)^{l_{ij}} | 0 \rangle ~~~ \nonumber 
\end{eqnarray}
In (\ref{fr}), the prime over the summation means that the 
linking numbers $l_{13},l_{14},l_{23},l_{24}$ are summed over all possible values which 
are consistent with (\ref{part}). More simply, in the loop network language, the 
summations are over all possible contractions of the abelian flux lines keeping $l_{12}$ 
contractions fixed \footnote{Our derivation of (\ref{fr}) in arbitrary d dimension 
(see section III) is by exploiting the properties of SU(2) coherent states and will 
be published elsewhere \cite{manu2}.}. The normalization constant 
$N(j) = N\big(j_1,j_2,j_{12}\big)N\big(j_{12},j_{3},j_{123}\big)N
\big(j_{123}(=j_4)),j_4,0\big)$, where  
$N(j_1,j_2,j_3) = \big((j_1-j_2+j_3)!(-j_1+j_2+j_3)!(j_1+j_2-j_3)!\big)^{\frac{1}{2}} 
\big(\frac{(2j_3+1)}{(j_1+j_2+j_3+1)!}\big)^{\frac{1}{2}}$.   
In (\ref{fr}), on the left hand site $j_3$ and $j_4$ are within 
brackets as they are associated with the previous sites due to 
the U(1) Gauss law (\ref{u1}). 
\begin{figure}[t]
\begin{center}
\includegraphics[width=0.39\textwidth,height=0.13\textwidth]
{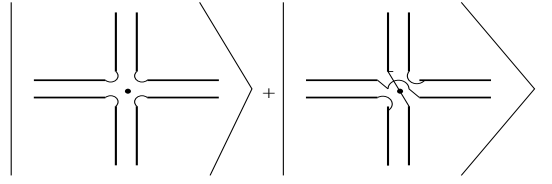}
\end{center}
\vspace{-5mm}
\caption{The loop state at site n $|j_{1}=j_{2}=j_{3}=j_{4}=j_{12} =1>$ 
from the orthonormal set (\ref{fr}) corresponding to Figure 1. The other  
state $|j_{1}=j_{2}=j_{3}=j_{4}=1, j_{12} =0>$ constructed from Figure 1 
and orthogonal to $|j_{1}=j_{2}=j_{3}=j_{4}=j_{12} =1>$ is 
represented by choosing the linking numbers $l_{12}=l_{34}=2$.}
\label{fig:figure2}
\end{figure}
Thus in d=2 the $SU(2)\otimes U(1)$ the gauge invariant 
loop network basis is characterized  by three physical 
(gauge invariant) quantum numbers per 
lattice site. We summarize the 
results obtained so far: In d = 2, the complete set of orthonormal gauge invariant 
states is isomorphic to the set of all possible loops which are labeled by 
the number of forward loop lines $j(n,1), j(n,2)$ and the linking number $l_{12}(n)$
at every lattice site.  The explicit construction is given by (\ref{fr}). 

\subsection{The Loop Space Dynamics} 

The Hamiltonian (\ref{ham}) has a very simple interpretation 
in the dual loop basis (\ref{fr}). The electric field term 
is now like potential energy term which simply counts the number 
of abelian flux lines. It's contribution to the energy is: 
$\sum_{links(l)} j(l)\left(j(l) +1\right)$. The plaquette term 
in (\ref{ham}) too has a simple meaning: it creates or 
annihilates the abelian flux lines on the plaquette. 
This can be seen by writing the link operator in terms of 
prepotentials: 
\begin{eqnarray}
U({\it l})_{\alpha\beta}  &=& F({\it l})  \big(a^{\dagger}_{\alpha}(l)
\tilde{b}^{\dagger}_{\beta}(l) +\tilde{a}_{\alpha}(l) 
{b}_{\beta}(l)\big) F(l) \nonumber \\
& \equiv &  U^{+}_{\alpha\beta}({\it l}) + U^{-}_{\alpha\beta}({\it l}).
\label{dhh}
\end{eqnarray}
Above, $F({\it l}) = 
\frac{1}{\sqrt{[a^{\dagger}({\it l}).a({\it l})+1]}}$.  
The transformation to prepotentials (\ref{dhh}) is obvious from the 
$SU(2) \otimes U(1)$ gauge transformations (\ref{gt3}) and (\ref{u1})
respectively. Looking at (\ref{dhh}) we realize that the operator 
$U^{+}({\it l})$ $(U^{-}({\it l}))$ creates (destroys) an abelian flux 
line on the link $({\it l})$ like in the case of compact QED 
(see also \cite{sharat2}).  We now compute the matrix elements of the 
Hamiltonian in the loop basis 
(\ref{fr}) for d=2. For convenience, we denote the four corners: 
n,n+1,n+1+2,n+2 of the plaquette 
located at n by a,b,c,d respectively and the associated loop 
basis vector as 
$|j_{abcd}> \equiv 
\prod_{x=a,b,c,d} \otimes |j_{1}^{x},j_{2}^{x},j_{12}^{x}>$. 
We consider the following $SU(2) \otimes U(1)$ gauge invariant part of 
the plaquette term in (\ref{ham}):   
$U_{plaq} = \left(a^{\dagger}[1].\tilde{a}^{\dagger}[2]\right)_{a}   
\left(a^{\dagger}[2].\tilde{a}^{\dagger}[3]\right)_{b}   
\left(a^{\dagger}[3].\tilde{a}^{\dagger}[4]\right)_{c}  
\left(a^{\dagger}[1].\tilde{a}^{\dagger}[4]\right)_{d}$ + h.c.   
The matrix elements are \cite{manu2}: 
\begin{eqnarray} 
&&<\bar{j}_{abcd}|U_{plaq}|j_{abcd}> =  
\Big(N_{+}~\delta_{\bar{j}_{1}^{a},{j}_{1}^{a}+\frac{1}{2}}  
\delta_{\bar{j}_{2}^{b},{j}_{2}^{b}+\frac{1}{2}}  
\delta_{\bar{j}_{1}^{d},{j}_{1}^{d}+\frac{1}{2}}  
\nonumber \\  
&& \delta_{\bar{j}_{2}^{a},{j}_{2}^{a}+\frac{1}{2}} +    
N_{-}~\delta_{{j}_{1}^{a},{j}_{1}^{a}-\frac{1}{2}}  
\delta_{\bar{j}_{2}^{b},{j}_{2}^{b}-\frac{1}{2}}  
\delta_{\bar{j}_{1}^{d},{j}_{1}^{d}-\frac{1}{2}}  
\delta_{\bar{j}_{2}^{a},{j}_{2}^{a}-\frac{1}{2}}\Big) \nonumber \\ 
&&\left\{ \begin{array}{cccc}
j_{12}^{b} & \bar{j}_{12}^{b} & \frac{1}{2}  \\
\bar{j}_{1}^{a} & j_{1}^{a}  & j_{4}^{b}\\
\end{array} \right \} 
\left\{ \begin{array}{cccc}
j_{12}^{b} & \bar{j}_{12}^{b} & \frac{1}{2}  \\
\bar{j}_{2}^{b} & j_{2}^{b}  & j_{1}^{b}\\
\end{array} \right \} 
\left\{ \begin{array}{cccc}
j_{12}^{d} & \bar{j}_{12}^{d} & \frac{1}{2}  \\
\bar{j}_{1}^{d} & j_{1}^{d}  & j_{2}^{d}\\
\end{array} \right \}  \nonumber \\
&& \left\{ \begin{array}{cccc}
j_{12}^{d} & \bar{j}_{12}^{d} & \frac{1}{2}  \\
\bar{j}_{2}^{a} & j_{2}^{a}  & j_{3}^{d}\\
\end{array} \right \}. ~~~
\label{me} 
\end{eqnarray} 
Above $N_{\pm}$ are the constants depending on the 
angular momentum quantum numbers on the plaquette 
$(abcd)$. The trivial $\delta$ functions over the 
quantum numbers which do not change are not shown.  
The 6-j symbols simply reflect the spin half 
nature of the prepotentials. The details will be given 
elsewhere \cite{manu2} (also see \cite{sharat1}).   

\section{The Loop States in d Dimension} 

It is easy to generalize d=2 construction of the previous section. 
We extend the angular momentum ladder and choose: $J[1]+J[2] \rightarrow J[12]+J[3] 
\rightarrow J[123] + ...\rightarrow J[12..2d-1]+J[2d] = J = 0$.  The states are now 
characterized as: $|j_1,j_2,..,j_d,(j_{d+1}),..,(j_{2d}),j_{12},j_{123},
..j_{12...2d-1}=(j_{2d})> \equiv |j_1,j_2,..,j_d,j_{12},
j_{123},..,j_{12...2d-2}>$. Thus the loop network  
is labeled by $3(d-1)$ gauge invariant 
angular momentum quantum numbers at every lattice 
site which is the number of transverse physical 
degrees of the freedom of the gluons \cite{sharat1}. 
The states are again given by (\ref{fr}) with  
the constraints on the linking numbers which have to be 
generalized. 
The Young tableau arguments like in d=2 case, lead to:  
$l_{12}=j_{1}+j_{2}-j_{12},l_{13}+l_{23}=j_{12}+j_{3}-j_{123}, 
......,l_{1,2d}+l_{2,2d}+.....l_{2d-1,2d} = 2j_{2d-1} 
\equiv 2j_{2d}$. 
Note that the last equation is an identity. 
As in the d=2 case, all possible contractions consistent with 
the above constraints and with the number of flux lines on the links 
(\ref{part}) are required to get the manifestly gauge invariant 
orthonormal loop basis. The dynamical matter fields  
are easy to incorporate, they will provide SU(2) charge sources and 
sinks to the abelian flux lines at their end points (see Figure 1) 
leading to additional color singlets. 

\section{SU(N) Lattice Gauge Theory} 

The SU(N) group has $(N-1)$ fundamental representations. Therefore, 
the defining equations for the SU(N) harmonic oscillator prepotentails 
\cite{manu1} on the links are: 
\begin{eqnarray}
E_{L}^{a}  \equiv  \sum_{r=1}^{(N-1)} a^{\dagger}[r]\frac{\lambda^{a}[r]}{2} a[r],  
E_{R}^{a}  \equiv   \sum_{r=1}^{(N-1)} b^{\dagger}[r]\frac{\lambda^{a}[r]}{2} b[r]  
\nonumber 
\end{eqnarray}
where the prepotential oscillators a and b are defined at the initial and the 
end point of the link $l$ respectively, the index r varies over the rank of the SU(N) 
group. Thus, the SU(N) lattice gauge theories in terms of prepotentials will 
have $SU(N) \otimes U(1)^{(N-1)}$ gauge invariance. This will lead to $(N-1)$ 
types of abelian flux lines in the SU(N) loop space. The role of $(N-1)$ 
abelian gauge groups in the confinement mechanism of SU(N) gauge theories 
has been emphasized by `t Hooft through the idea of abelian projection \cite{hooft}. 
It is interesting to imagine the background SU(2) quark anti-quark pair 
located at two different lattice sites in the prepotential formulation. 
The SU(2) fluxes will be neutralized {\it locally} but the abelian 
gauge invariance will demand the formation of an abelian flux line (string)
between quark anti-quark pair leading to color confinement in the strong 
coupling limit.  The  construction of SU(N) loop basis, the issue 
of color confinement and especially $N \rightarrow \infty$ limit are under 
investigation. 

I would like to thank Rajiv Gavai and Sourendu Gupta for the 
hospitality at TIFR, Mumbai where part of the paper was written.


\begin{thebibliography}{99}

\bibitem{mans} S. Mandelstam, Phys. Rev. 175 (1968) 1580, A. M. Polyakov, 
Phys. Lett. {\bf B 82} (1979) 247, 
A. A. Migdal, Phys. Rep. {\bf 102} (1983) 199, 
S. Mandelstam, Phys. Rev. {\bf D 19} (1979) 2391,  
A.  M. Polyakov, Gauge Fields and Strings (Harwood, New York, 1987). 
\bibitem{ash} A. Ashtekar, Phys. Rev. Letts. {\bf 57} (1986) 2244, C. Rovelli, L. Smolin, 
Phys Rev. {\bf D 52} (1995) 5743.  
\bibitem{rst} Br\"ugmann B  1991 Phys. Rev. D {\bf 43} 566, Loll R 1992 Nucl. 
Phys B 368 121, Watson N J 1994 Phys. Letts. B 323 385, 
R. Gambini, Lorenzo Leal, Antoni Trias, Phys. Rev. {\bf D 39} (1989) 3127, 
Bartolo C,  Gambini R, Leal L 1989 Phys. Rev. {\bf D 39} 1756. 
\bibitem{manu} M. Mathur, to be published in Journal of Physics A.   
\bibitem{sharat1} R. Anishetty, H. S. Sharatchandra, Phys. Rev. Letts. {\bf 65} (1990) 81. 
H. S. Sharatchandra, Nuclear Physics  {\bf B 196} (1982) 62. 
\bibitem{sharat2} B. Gnanapragasam, H. S. Sharatchandra, Phys. Rev. {\bf D 45} 
(1992) R1010. 
\bibitem{pol} A. M. Polyakov, Phys. Letts. {\bf B 59}, (1975) 82, T. Banks, R. Myerson, 
J. Kogut, Nucl. Phys. {\bf B129} (1977) 473. 
\bibitem{kogut} J. Kogut, L Susskind, Phys Rev. {\bf D 10} (1974) 3468.  
\bibitem{schwinger} J. Schwinger 1952 D. Mattis, {\it The Theory of Magnetism} (Harper and Row, 1982). 
\bibitem{manu2} Manu Mathur, under preparation. 
\bibitem{rob} D. Robson, D. M. Weber, Z. Phys. {\bf C15} (1982), 199. \\ 
W. Furmanski, A. Kolawa, Nucl. Phys., {\bf B 291}, (1987) 594  
G. Burgio, R. De. Pietri, H. A. Morales-Tecotl, L. F. Urrutia, 
J. D. Vergara, Nucl. Phys. {\bf B 566}, (2000), 547.  
\bibitem{manu1} M. Mathur and D. Sen, J. Math. Phys. {\bf 42} (2001) 4181, 
M. Mathur, H. S. Mani, J. Math. Phys. {\bf 43} (2002) 5351.
\bibitem{hooft} G. `t Hooft,  Nucl.  Phys. {\bf B 190} [FS 3] (1981) 455.  
\end{thebibliography}
\end{document}